\documentclass[prb,aps,twocolumn,showpacs,preprintnumbers,amsmath,amssymb]{revtex4}
\usepackage{graphicx}
\usepackage{dcolumn}

\begin{document}
\title{Antiferromagnetic and structural transitions in the superoxide KO$_{2}$
    from first principles: A 2$p$-electron system with spin-orbital-lattice
    coupling}
\author{Minjae Kim, Beom Hyun Kim, Hong Chul Choi, and B. I. Min}
\affiliation{Department of Physics, PCTP,
	Pohang University of Science and Technology, Pohang, 790-784, Korea}
\date{\today}

\begin{abstract}
KO$_{2}$ exhibits concomitant antiferromagnetic (AFM) and
structural transitions, both of which originate from
the open-shell $2p$ electrons of O$_{2}^{-}$ molecules.
The structural transition is accompanied by the coherent tilting of
O$_{2}^{-}$ molecular axes.  The interplay among the
spin-orbital-lattice degrees of freedom in KO$_{2}$
is investigated by employing the first-principles electronic
structure theory and the kinetic-exchange interaction scheme.
We have shown that the insulating nature of the high symmetry phase
of KO$_{2}$ at high temperature (T) arises from the combined effect of
the spin-orbit coupling and the strong Coulomb correlation of
O 2$p$ electrons.  In contrast, for the low symmetry phase of KO$_{2}$
at low T with the tilted O$_{2}^{-}$ molecular axes,
the band gap and the orbital ordering are driven by
the combined effects of the crystal-field and the strong
Coulomb correlation.
We have verified that the emergence of the O 2$p$ ferro-orbital ordering
is essential to achieve the observed AFM structure for KO$_{2}$.
\end{abstract}

\pacs{75.50.Ee, 71.70.Ej, 71.15.Mb}

\maketitle

Magnetism due to the correlated 2$p$ electrons has attracted revived
attention for the possibility of the new kinds of the magnetic
informative materials.\cite{Attema07,Solovyev}
Some 2$p$ magnetic oxides exhibit the structural
phase transition concomitantly with the magnetic phase
transition.\cite{Meier,Kanzig,Labhart}
Solid oxygen is a typical example, which has both
antiferromagnetic (AFM) and structural phase transitions
below 24 K.\cite{Meier}
Alkali superoxides, $A$O$_{2}$ ($A$=Na, K, Rb), which are of our
present interest, belong to another example.\cite{Hesse89,Solovyev}
In $A$O$_{2}$, one alkali-metal atom provides one electron
to an oxygen molecule, and thereby each O$_{2}^{-}$ anion has
nine electrons at the 2$p$ molecular levels with
the electronic configuration of $\sigma_{g}^{2}\pi_{u}^{4}\pi_{g}^{3}$
(see the inset of Fig.~\ref{fig2}).\cite{Attema07,Attema05}
The partially occupied antibonding $\pi_{g}$ molecular states
play the most important role in determining the electronic and magnetic
properties of alkali superoxides.\cite{Bosch}
One hole in $\pi_{g}$ generates the  magnetic moment of
1 $\mu_{B}$ for each O$_{2}^{-}$.
The degeneracy of the $\pi_{g}$ level is expected to be lifted
by lowering the crystal symmetry,
as occurs due to the Jahn-Teller effect.\cite{Bosch}
In fact, it was suggested
that, for KO$_{2}$, the symmetry lowering would occur
$via$ coherent tilting of the O$_{2}^{-}$ molecular axes,
the so-called magnetogyration, which invokes the accompanying AFM
ordering.\cite{Kanzig,Labhart,Lines}

\begin{figure}[b]
\includegraphics[width=8cm]{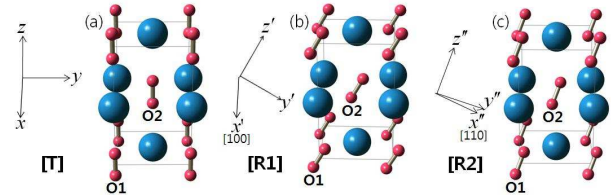}
\caption{(Color online)
(a) The tetragonal [T] structure of KO$_{2}$
with unrotated O$_{2}^{-}$ molecular bond axes.
The molecular axes are along the $z$-direction.
K atoms are in blue and O atoms in red.
(b) [R1] with molecular bonds rotated by $\sim 30^{\circ}$
around the [100] ($x'$) axis with
bond axes along the $z'$-direction.
(c) [R2] with molecular bonds rotated by $\sim 30^{\circ}$ around
the [110] ($x''$) axis with bond axes along the $z''$-direction.
There are two independent types of oxygen O1 and O2
for each structure.
}
\label{fig1}
\end{figure}

\begin{figure}[t]
\includegraphics[width=8cm]{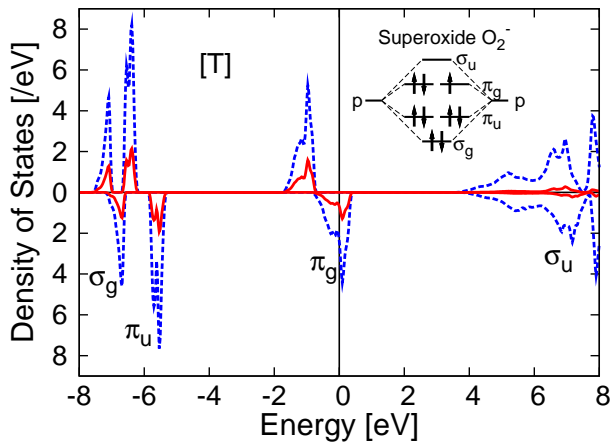}
\caption{(Color online) The DOS in the GGA for the FM phase of
KO$_{2}$ with [T] structure.
Dotted (blue) and solid (red) lines represent the total DOS
and the partial DOS of oxygen $2p$ states, respectively.
The upper and the lower panels correspond to the majority and minority
spin DOSs. Inset shows the 2$p$ molecular levels of O$_{2}^{-}$.
}
\label{fig2}
\end{figure}

At room temperature, KO$_{2}$ crystallizes in
the tetragonal structure of CaC$_{2}$ type,
in which the O$_{2}^{-}$ molecular bond axes are parallel to
the $z$-axis (Fig.~\ref{fig1}(a)).\cite{Kanzig}
KO$_{2}$ retains this structure down to 197 K and
exhibits the paramagnetic behavior.
Upon cooling, O$_{2}^{-}$ molecular bond axes seem to
tilt uniformly by $\sim 20 ^{\circ}$ to have a lower
crystal (monoclinic) symmetry.
The magnetic phase is still paramagnetic down to 7 K.
Below 7 K, the AFM ordering emerges in the triclinic crystal structure
with the uniform tilting of O$_{2}^{-}$ molecular bond axes
by $\sim 30^{\circ}$.\cite{Kanzig}
According to neutron experiment,\cite{Smith} the AFM phase
has the magnetic structure having opposite spin arrangements
along the $z$-direction between two oxygen layers of O1 and O2.
This feature in KO$_{2}$ reflects the strong interplay among
spin, orbital, and lattice degrees of freedom,
as in rare-earth manganites.\cite{Solovyev}

There have been several theoretical reports to study the
coupled structural and magnetic transitions in
KO$_{2}$,\cite{Bosch,Lines,Solovyev,Kemeny,Lines2}
but those studies were mostly qualitative
and lacked the quantitative description of the electronic structures
for the low symmetry phase of KO$_{2}$.
Even the direction of the tilted molecular bond axis
is still uncertain between two possibilities.
The first one is [R1] in Fig.~\ref{fig1}(b),
in which the molecular bonds are rotated around the [100] axis.
The second one is [R2] in Fig.~\ref{fig1}(c),
in which the molecular bonds are rotated around [110] axis.
Moreover, the recent band structure study for the high symmetry
phase of KO$_{2}$ ((Fig.~\ref{fig1}(a)) in the local density
approximation (LDA)\cite{Solovyev} reveals that
the degenerate $\pi_{g}$ states do not
split, resulting in the metallic nature,
which is contradictory to the insulating nature of KO$_{2}$.

In this Rapid Communication, we have investigated comprehensively
the electronic structures of both high and low symmetry phases
of KO$_{2}$, and explored the mechanism of the
interplay between the spin and lattice degrees of freedom in
correlated $2p$ electron systems. Further, we have examined
the origins of the band gap opening and the exchange interactions
between  O$_{2}^{-}$ molecules in KO$_{2}$.

We have employed the full-potential augmented plane
wave (FLAPW) band method\cite{FLAPW}
implemented in WIEN2k package.\cite{Blaha}
For the exchange-correlation potential, the generalized gradient
approximation (GGA)\cite{GGA} was used.
We also incorporate the on-site Coulomb interaction $U$
between the oxygen $2p$ electrons (GGA+$U$)
and the spin-orbit (SO) effect as a second variational procedure
(GGA+$U$+SO).
The valence wave functions inside the muffin-tin spheres are expanded
in spherical harmonics up to $l_{max}$=10, and the wave function
in the interstitial region is expanded with plane waves up to
$K_{max}=7/R_{\textrm{MT}}$,
where $R_{\textrm{MT}}$ is the smallest muffin-tin sphere radius.
$R_{\textrm{MT}}$'s were set as 2.00 (a.u.) for K and 1.23 (a.u.)
for Oxygen, respectively.
The charge density was expanded  with plane waves up to
$G_{max}$=12 (a.u.)$^{-1}$.
We have used 200 k-points inside the first Brillouin zone.
For [R1] and [R2] in Fig.~\ref{fig1},
the O$_{2}^{-}$ molecular bonds were assumed to
be rotated uniformly by 30$^{\circ}$ from [T] structure
around the [100] and [110] axis, respectively.
The experimental lattice parameters $a$=4.030{\AA} and $c$=6.697{\AA}
and the bond length $d$$_{oo}$=1.306{\AA} were used.\cite{tetra}
The spin direction was chosen to be perpendicular
to the molecular axis following the experiment.\cite{Labhart}


Figure~\ref{fig2} provides the density of states (DOS)
in the GGA band calculation for the FM phase of KO$_{2}$
with the high symmetry un-rotated structure [T].
Oxygen 2$p$ partial DOS (PDOS)
shows clear molecular level splittings among $\sigma_{g},
\pi_{u}, \pi_{g}$ and $\sigma_{u}$ in agreement with
literature.\cite{Solovyev}
The finite DOS at the Fermi level E$_F$ in the minority spin $\pi_{g}$
states produces the half-metallic character for KO$_{2}$,
which reflects the failure of the GGA.
The degenerate $\pi_{g}$ states, which are composed of anti-bonding
$\pi_{x}$ and $\pi_{y}$ states, should split for an insulating
phase of KO$_{2}$.
To understand the splitting mechanism of the $\pi_{g}$ states,
we have considered the effects of the SO and the
Coulomb correlation of O 2$p$ electrons.
The importance of the Coulomb correlation effect of O 2$p$ electrons
was examined recently in a similar $2p$ magnetic oxide
Rb$_{4}$O$_{6}$.\cite{Winterlik07,Winterlik}

\begin{figure}[t]
\includegraphics[width=8cm]{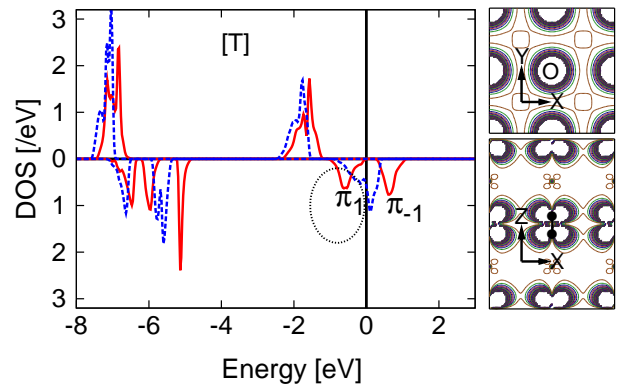}
\caption{(Color online) The PDOS of oxygen $2p$ states
for the FM phase of KO$_{2}$ with [T] structure.
The dotted (blue) and solid (red) lines represent the DOSs
in the GGA+$U$ and the GGA+$U$+SO, respectively ($U$=6.53 eV).
The $\pi_{g}$ states split into $\pi_{m=1}, \pi_{m=-1}$ in
the GGA+$U$+SO. Charge densities of $\pi_{m=1}$ states are plotted
on the $xy$ (001) (the upper panel) and on the $xz$ (010) plane
(the lower panel).
}
\label{fig3}
\end{figure}

Figure~\ref{fig3} shows the DOS both
in the GGA+$U$ and GGA+$U$+SO for the FM phase of KO$_{2}$
with [T] structure.
Note that the $\pi_{g}$ states near E$_F$ do not split
in the GGA+$U$ with $U$=6.53 eV.
They become split only when the SO is included in the GGA+$U$:
$\pi_{g}$ split into $\pi_{m=1}$
and $\pi_{m=-1}$ which are mixed states of $\pi_{x}$ and $\pi_{y}$.
In the right of Fig.~\ref{fig3}, the charge densities for the
occupied $\pi_{m=1}$ states are plotted on the (001) and (010) planes.
These charge densities manifest the azimuthally symmetric nature
of $\pi_{m=1}$ states, which
demonstrates that the splitting of $\pi_{g}$ states originates from
the SO effect. In fact, the large SO effect results from
the Coulomb correlation, which localizes the 2$p$ electrons
to generate the substantial orbital magnetic moment in the
$\pi_{g}$ states. The magnitude of the orbital magnetic moment is
as much as 0.570 $\mu_{B}$
per O$_{2}^{-}$ molecule, which is comparable to that of the
spin magnetic moment of 1.0 $\mu_{B}$.
The direction of the orbital magnetic moment turns out to be
almost parallel to the molecular axis, {\it i.e.}, perpendicular to
that of the spin magnetic moment.
We have checked that the electronic structure of AFM KO$_{2}$ with
[T] structure is close to that of FM KO$_{2}$, and so
the band gap at E$_F$ opens in the same way.

\begin{figure}[t]
\includegraphics[width=8.2cm]{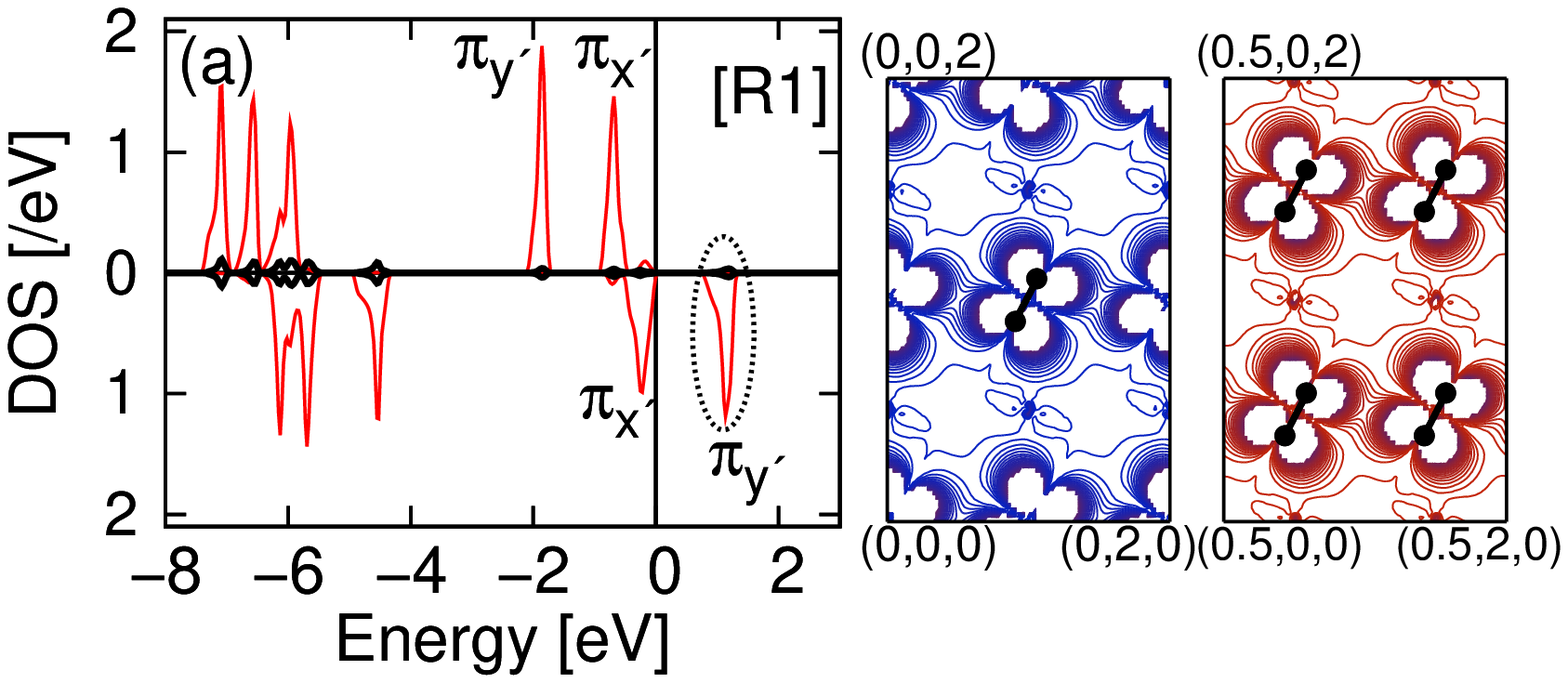}\\
\includegraphics[width=8.2cm]{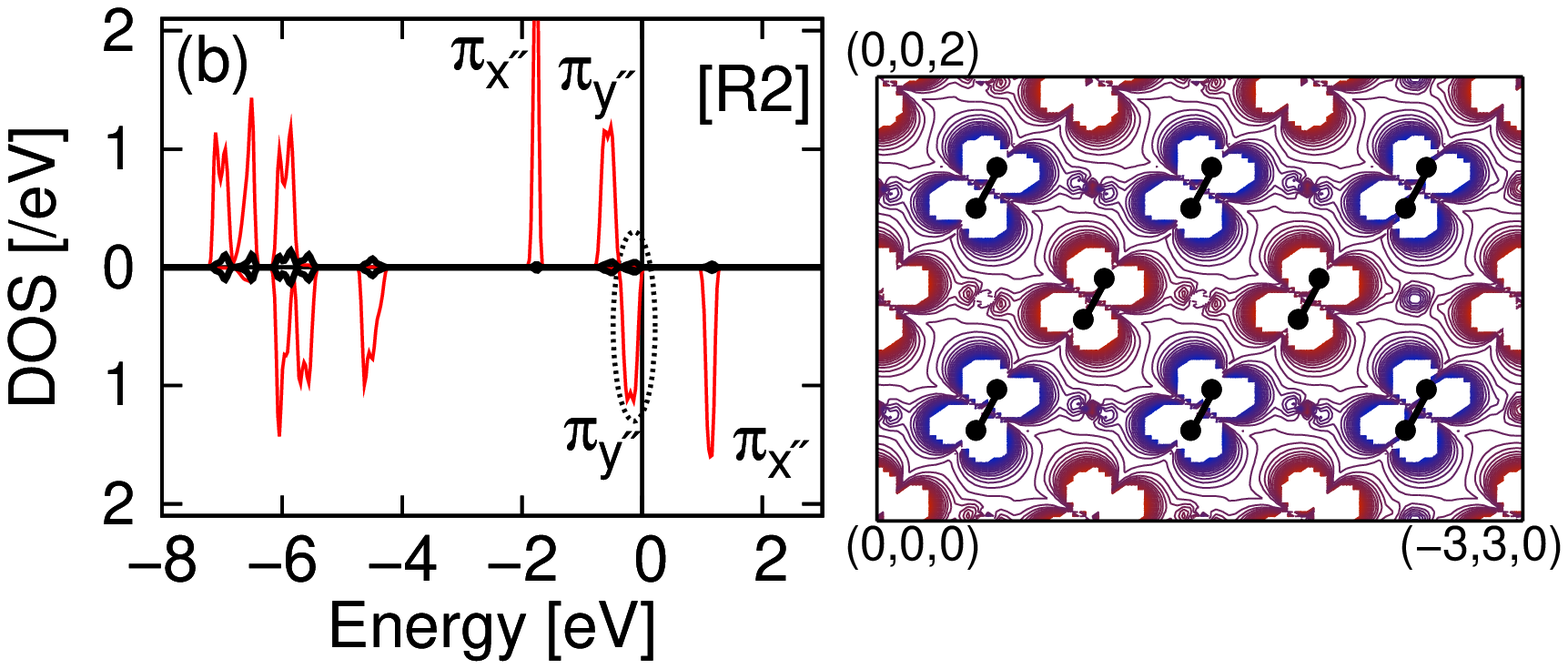}\\
\caption{(Color online)
(a) The local PDOS of the O $2p$ states in the GGA+$U$+SO ($U$=6.53 eV)
  for the AFM phase of KO$_{2}$ with [R1] structure.
The spin densities of unoccupied $\pi_{y'}$ states are plotted on
the (100) and (200) planes. Blue and red colors in the spin densities
represent opposite spins.
(b) The local PDOS of the O $2p$ states in the GGA+$U$+SO ($U$=6.53 eV)
  for the AFM phase of KO$_{2}$ with [R2] structure.
  The spin density of occupied minority spin $\pi_{y''}$ states
  is plotted on the $(110)$ plane.
Black solid lines in the DOS plots represent the PDOS of K $4p$ states.
}
\label{fig4}
\end{figure}
\begin{table}[!b]
\caption{Magnetic exchange constants [meV] between O$_{2}^{-}$ molecules
for direct and indirect ($via$ K$^{+}$) hopping channels
in the [R1] and [R2] structures of KO$_{2}$.
$J_{1}$ and $J_{2}$ are the in-plane exchange constants
along the $x$ and the $y$-direction,
and $J_{3}$ is the inter-plane exchange constant.
Positive $J$ represents the AFM interaction.
}
\begin{tabular}{l  c  r @{.} l r @{.} l  r @{.} l}
\hline \hline
		&~	& \multicolumn{2}{c}{$J_{1}$}
                        & \multicolumn{2}{c}{$J_{2}$}
                        & \multicolumn{2}{c}{$J_{3}$} \\ \hline
R1 (direct)  	&~	& ~0&353 & ~1&486 & ~~0&209 \\
R1 ($via$ K$^{+}$)  &~	& ~0&023 & ~0&023 & ~~0&027 \\
R2 (direct)  	&~	& ~0&419 & ~0&419 & ~~0&680\\
R2 ($via$ K$^{+}$)  &~	& $-$0&215 & $-$0&215 & ~~0&005\\
\hline\hline
\end{tabular}
\label{table1}
\end{table}

\begin{figure}[t]
\includegraphics[width=6.5cm]{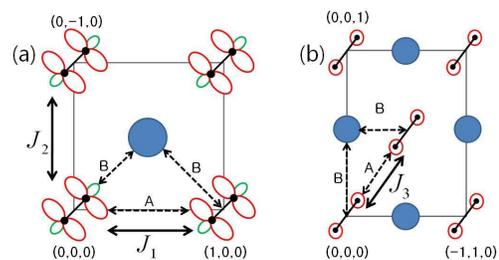}
\caption{(Color online)
(a) The in-plane exchange constants, $J_{1}$ and $J_{2}$,
on the $xy$-plane for KO$_{2}$ with [R2],
where the unoccupied $\pi_{x''}$ and occupied $\pi_{y''}$ states
are represented by red and green, respectively.
(b) The inter-plane exchange constant $J_{3}$ on the (110) plane.
The paths A and B correspond to the direct and indirect ($via$ K$^{+}$)
hopping channels, respectively.
}
\label{fig5}
\end{figure}

For KO$_{2}$ with [R1] and [R2] structures,
the crystal field effect will be activated due to tilting
of O$_{2}^{-}$ molecular axes toward K$^{+}$.
Figure~\ref{fig4}(a) shows the local PDOSs and the spin densities
in the GGA+$U$+SO for the AFM phase of KO$_{2}$ with [R1] structure.
Notable feature in the DOS is the opening of band gap at E$_F$.
The GGA band calculation does not produce
the band gap, whereas both the GGA+$U$ and the GGA+$U$+SO yield
almost the same insulating band structure.
This implies that, in producing the band gap,
(i) the symmetry lowering just by tilting of molecular axes
is not sufficient, (ii) the Coulomb correlation effect is essential, and
(iii) the SO effect for [R1] is not as important as that for [T].
The orbital magnetic moment is almost quenched having
only 0.002 $\mu_{B}$ per O$_{2}^{-}$ molecule.
Therefore, it is the combined effect of the Coulomb $U$ and
the crystal field from K$^{+}$ cations that
splits the degenerate $\pi_{g}$ states into the directional orbitals,
$\pi_{x'}$ and $\pi_{y'}$ states.
Interestingly, due to the complex interplay of the Coulomb $U$ and
the crystal field effect in the molecular states,
the splitting of $\pi_{g}$ states occurs extraordinarily,
{\it i.e.}, the $\pi_{y'}$-$\pi_{x'}$ order for the majority spin
and the opposite $\pi_{x'}$-$\pi_{y'}$ order for the minority spin states.
The half-filled molecular states $\pi_{y'}$ seem to be strongly affected
by Coulomb $U$, while the fully occupied molecular states $\pi_{x'}$
seem to be weakly affected.
The spin densities of unoccupied $\pi_{g}$ states plotted on the (100)
and (200) planes demonstrate that they really correspond
to the $\pi_{y'}$ states. The $\pi_{y'}$ states in each plane
exhibit the FM spin and ferro-orbital (FO) orderings, while
they exhibit the inter-plane AFM spin ordering.
Namely, the AFM and FO orderings take place concurrently.

Similarly, Fig.~\ref{fig4}(b) shows the local PDOSs
and the spin densities
in the GGA+$U$+SO for AFM KO$_{2}$ with [R2] structure.
For [R2] too, the GGA does not produce the band gap,
whereas both the GGA+$U$ and the GGA+$U$+SO produce it.
The SO effect is again negligible
so that the orbital magnetic moment is quenched having only
0.008 $\mu_{B}$ per O$_{2}^{-}$ molecule.
Degenerate $\pi_{g}$ states are split into $\pi_{x''}$ and $\pi_{y''}$
due to the combined effect of the Coulomb $U$ and the crystal field
from K$^{+}$.
The spin density of occupied minority spin $\pi_{g}$ states
plotted on the $(110)$ plane demonstrates that they
really correspond to the $\pi_{y''}$ states.

Now let us examine the magnetic interaction in KO$_{2}$
based on the above electronic structures.
The implication of Fig.~\ref{fig4} is that,
with the tilting of O$_{2}^{-}$ molecular axes in the AFM
[R1] and [R2] structures, the band gap opening and the FO
ordering in $\pi_{g}$ states occur simultaneously.
According to the spin-orbital model for 3$d$ transition-metal (TM)
oxides,\cite{SX}
the FO ordering of TM 3$d$ states would induce the AFM superexchange
interaction between two TM spins, and vice versa.
In this respect, the results in Fig.~\ref{fig4} look reasonable.
The superexchange, more generally, the kinetic exchange in KO$_{2}$
will take place through two channels:
the direct hopping between two O$_{2}^{-}$ molecules
and the indirect hopping $via$ K$^{+}$ (see Fig.~\ref{fig5}).
The unoccupied K $4p$ states are located $4-6$ eV above E$_F$ (see
Fig.~\ref{fig2}), and so there is a possibility of hopping mechanism
through O$_{2}^{-}$-K$^{+}$-O$_{2}^{-}$.
Indeed, as revealed in Fig.~\ref{fig4}, there exists
the nonnegligible hybridization between the O $2p$ and K $4p$ states.

Table~\ref{table1} presents the estimated exchange constants based on the
microscopic calculation of the kinetic exchange interaction
for KO$_{2}$ with the [R1] and [R2] structures.\cite{Bhkim}
The 2$p$ electrons in each O$_{2}^{-}$ are assumed to have the
atomic orbital states as in our band results for [R1] and [R2].
The kinetic exchange interactions both from the direct and indirect
hoppings were considered independently to calculate
the exchange constant for each channel.
For [R1], the exchange constants are obtained to be all AFM,
and the dominant channel is the direct hopping between
O$_{2}^{-}$ anions along the $y$-direction,
which is AFM ($J_{2}$=1.486 meV).
The large $J_{2}$ results from the tilting of
O$_{2}^{-}$ molecular axis along the [010].
The inter-plane interactions from both direct and indirect channels
yield smaller exchange constants ($J_{3}$=0.209 and 0.027 meV).
Then the resulting magnetic structure for [R1] will not
be consistent with the experimental AFM structure which has
alternating AFM spins along [001].

On the other hand, the kinetic exchange interactions for [R2]
turn out to be consistent with the experimental AFM structure.
The dominant one is the inter-plane AFM
interaction coming from the direct hopping ($J_{3}$=0.680 meV).
The in-plane exchange constants are 0.419 and $-0.215$ meV
for direct and indirect channels, respectively.
Hence the total exchange constants between two nearest neighbor
(NN) O$_{2}^{-}$ along $x$ and $y$-directions are 0.204 meV.
For each O$_{2}^{-}$ in [R2], there are eight
inter-plane NN O$_{2}^{-}$ and four in-plane NN O$_{2}^{-}$.
As a consequence, much larger $J_{3}$ than $J_{1}$ and $J_{2}$
will generate the interlayer AFM ordering along [001],
as is consistent with the experimental AFM structure.

The in-plane FM interactions ($-0.215$ meV)
coming from the indirect hopping $via$ K$^{+}$
play an important role in stabilizing the in-pane FM ordering.
As shown in Fig.~\ref{fig5}(a),
out of two neighboring O$_{2}^{-}$ molecules on the
$xy$ plane in [R2], the lobes of $\pi_{x''}$ orbital of one molecule
are toward the intermediate K$^{+}$, while those of the other molecule
are away from K$^{+}$ and thereby the lobes of $\pi_{y''}$ orbital
are toward the intermediate K$^{+}$.
This arrangement of the orbitals suggests that
the hopping takes place between the occupied $\pi_{y''}$ and
unoccupied $\pi_{x''}$ orbitals of two neighboring O$_{2}^{-}$ molecules.
This is reminiscent of the resulting FM superexchange interaction
between the fully-occupied and half-filled orbitals in TM oxides
in the framework of the Goodenough-Kanamori-Anderson (GKA) rule.\cite{GKA}
In contrast, for the inter-plane kinetic exchange in Fig.~\ref{fig5}(b),
the lobes of $\pi_{x''}$ orbitals of both O$_{2}^{-}$ molecules
are away from  K$^{+}$, and so
the indirect kinetic exchange results in the weak AFM interaction
($J_{3}$=0.005 meV).


In conclusion, we have investigated electronic and magnetic structures
of KO$_{2}$ superoxide, the strongly correlated 2$p$ electron system.
We have found that, for the correct description of the insulating
electronic structure of KO$_{2}$, the SO coupling as well as
the large Coulomb correlation is important for the high symmetry phase,
while, for the low symmetry phase,
the crystal field from K$^{+}$, as well as the large Coulomb
correlation, is important.
The concurrent AFM spin and FO orderings with the band-gap opening
clearly demonstrate the strong coupling among
the spin-orbital-lattice degrees of freedom in KO$_{2}$.
In the low symmetry phase of KO$_{2}$ with [R2] structure,
the emergent FO ordering yields the kinetic exchange interactions
that are consistent with the experimental AFM structure.

This work was supported by the NRF (No.2009-0079947),
and by the POSTECH Research Fund. Helpful discussions
with Y. H. Jeong and J.-S. Kang  are greatly appreciated.


\begin{thebibliography}{99}
\bibitem{Attema07} J. J. Attema, G. A. de Wijs, and R. A. de Groot,
	J. Phys.: Condens. Matter, \textbf{19}, 165203 (2007).
\bibitem{Solovyev}{I. V. Solovyev,
	New J. Phys. \textbf{10}, 013035 (2008).}
\bibitem{Meier}{R. J. Meier and R. B. Helmholdt,
	Phys. Rev. B \textbf{29}, 1387 (1984).}
\bibitem{Kanzig}{W. K\"{a}nzig and M. Labhart, 
	J. Phys. (paris) \textbf{37}, C7-39 (1976).}
\bibitem{Labhart}{M. Labhart, D. Raoux, W. K\"{a}nzig, 
	 and M. A. B\"{o}sch, Phys. Rev. B \textbf{20}, 53 (1979).}
\bibitem{Hesse89} W. Hesse, M. Jansen, and W. Schnick, 
	Prog. Solid State Chem. \textbf{19}, 47 (1989).
\bibitem{Attema05} J. J. Attema, G. A. de Wijs, G. R. Blake, and 
	R. A. de Groot, J. Am. Chem. Soc. \textbf{127}, 16325 (2005).
\bibitem{Bosch} {M. A. B\"{o}sch, M. E. Lines, and M. Labhart, 
	Phys. Rev. Lett. \textbf{45}, 140 (1980).}
\bibitem{Lines} {M. E. Lines and M. A. B\"{o}sch,
	Phys. Rev. B \textbf{23}, 263 (1981).}
\bibitem{Smith}{H. G. Smith, R. M. Nicklow, L. J. Raubenheimer, and
	M. K. Wilkinson, J. Appl. Phys. \textbf{37} 1047 (1966)}
\bibitem{Kemeny} {G. Kemeny, T. A. Kaplan, S. D. Mahanti, and D. Sahu, 
	Phys. Rev. B \textbf{24}, 5222 (1981).}
\bibitem{Lines2} {M. E. Lines, Phys. Rev. B \textbf{24}, 5248 (1981).} 
\bibitem{FLAPW} M. Weinert, E. Wimmer, and A. J. Freeman,
	Phys. Rev. B \textbf{26}, 4571(1982).
\bibitem{Blaha}{P. Blaha, K. Schwarz, G.K.H. Madsen, D. Kavasnicka,
    J. Luitz, WIEN2k  (Karlheinz Schwarz, Technische Universitat Wien,
         Austria, 2001).}
\bibitem{GGA}{J. P. Perdew, K. Burke and M. Ernzerhof,
	Phys. Rev. Lett. \textbf{77}, 3865 (1996).}
\bibitem{tetra} {
        Since the triclinic deformation is very small for [R1] and [R2],
	we assumed the tetragonal structure in the band calculations.
	See Ref.\cite{Kanzig}.}
\bibitem{Winterlik07} J. Winterlik, G. H. Fecher, and C. Felser,
	J. Am. Chem. Soc., \textbf{129}, 6990 (2007).
\bibitem{Winterlik} {J. Winterlik, G. H. Fecher, C. A. Jenkins, C. Felser,
	C. M\"{u}hle, K. Doll, M. Jansen, L. M. Sandratskii,
	and J. K\"{u}bler, Phys. Rev. Lett. \textbf{102}, 016401 (2009).}
\bibitem{SX} {K. I. Kugel and D. I. Khomskii, Sov. Phys. Usp. \textbf{25},
	231 (1982).}
\bibitem{Bhkim} {B. H. Kim, Ph.D. Thesis,(POSTECH 2009).}
\bibitem{GKA}{J. B. Goodenough, {\it Magnetism and the Chemical Bond},
	Interscience Publ., N.Y.-lnd., (1976) }


\end{thebibliography}
\end{document}